\documentstyle[graphicx, times]{mn2e}

\title[Iron K-shell absorption is not from the WHIM]
{On why the Iron K-shell absorption in AGN is not a signature of the
local Warm/Hot Intergalactic Medium}

\newcommand{\PG}{PG~1211+143}

\newcommand{\arcm}{\hbox{$^\prime$}}

\newcommand{\ltsima}{$\; \buildrel < \over \sim \;$}
\newcommand{\simlt}{\lower.5ex\hbox{\ltsima}} 
\newcommand{\gtsima}{$\; \buildrel > \over \sim \;$}
\newcommand{\simgt}{\lower.5ex\hbox{\gtsima}} 

\author[J. Reeves et al.] {
James Reeves$^1$, Chris Done$^2$, Ken Pounds$^3$, Yuichi Terashima$^4$, Kiyoshi Hayashida$^5$
\and Naohisa Anabuki$^{5}$, Masahiro Uchino$^{5}$, Martin Turner$^{3}$
\\
$^1$Astrophysics Group, School of Physical and Geographical Sciences, Keele 
University, Keele, Staffordshire, ST5 8EH, UK.\\
$^2$Department of Physics, University of Durham, South Road, Durham DH1 3LE,UK\\
$^3$Department of Physics and Astronomy, University of Leicester, 
University Road, Leicester, LE1 7RH, UK.\\
$^4$Department of Physics, Ehime University, Matsuyama 790-8577, Japan\\
$^5$Department of Earth and Space Science, Osaka University,
Toyonaka, Osaka 560-0043, Japan
}

\date{Submitted to MNRAS}
\pagerange{\pageref{firstpage}--\pageref{lastpage}} \pubyear{2003}

\begin{document}

\def\aap{A\&A}
\def\apj{ApJ}
\def\apjl{ApJ}
\def\mnras{MNRAS}

\maketitle

\label{firstpage}

\begin{abstract}
We present a comparison between the 2001 XMM-Newton and 2005 Suzaku observations 
of the quasar, PG\,1211+143 at $z=0.0809$. Variability is observed 
in the 7 keV iron K-shell absorption line (at 7.6 keV in the quasar frame), which is 
significantly weaker in 2005 than during the 2001 XMM-Newton observation.    
From a recombination timescale of $<4$ years, this implies an absorber density  
$n>4\times10^{3}$\,cm$^{-3}$, while the absorber column 
is $5\times10^{22}<N_{\rm H}<1\times10^{24}$\,cm$^{-2}$. 
Thus the sizescale of the absorber is too compact (pc scale) and 
the surface brightness of the dense gas too high (by 9-10 orders of 
magnitude) to arise from local hot gas, such as the local bubble, group or 
Warm/Hot Intergalactic Medium (WHIM), as
suggested by McKernan et al. (2004, 2005). Instead the iron K-shell 
absorption must be associated with an AGN outflow with mildly relativistic 
velocities. Finally we show that the 
the association of the absorption in PG\,1211+143 with local hot gas is simply 
a coincidence, the comparison between the recession and iron K absorber outflow velocities 
in other AGN does not reveal a one to one kinematic correlation. 
\end{abstract}


\section{Introduction}

The cosmological requirement that half the baryons in the Universe are
in a warm/hot intergalactic medium (WHIM) has motivated UV and X--ray
absorption line studies to detect this otherwise invisible gas, using
absorption against a bright AGN to probe the line of sight material
(Bregman 2007). However, most AGN (except
blazars) have intrinsic columns of warm absorbing gas in their nuclei,
and the fact that this is generally connected to a nuclear 
outflow (Blustin et al 2005) complicates separating this
intrinsic absorption from extrinsic line--of--sight material using
velocity information. Nonetheless, various studies have looked for
absorption lines (predominantly OVII/VIII) as signatures of the WHIM,
though all current detections are more likely to be associated with
material in our Galactic halo or Local Group (Bregman \& Lloyd--Jones
2007). These have velocities within a few hundred km/s of the local
standard of rest, and indicate columns of OVII/VIII $>
10^{16}$~cm$^{-2}$, equivalent to Hydrogen columns of 
$N_H\sim 10^{19}$~cm$^{-2}$.

Much more controversial was the potential association of substantially
higher columns of gas to such local material. McKernan et al (2004;
2005) noticed that several AGN with strong He-- or H--like Fe
absorption features had these lines at energies which were
approximately consistent with the local standard of rest.  For any
single object the approximate match between the putative blueshifted
outflow velocity and the galaxy redshift could be coincidental, but
McKernan et al (2004; 2005) pointed out three AGN showing this trend,
with redshifts spanning $\sim 0.008-0.15$ (MCG--6--30--15, PG1211+143
and PDS~456). However, the derived columns of $N_H\sim
10^{23}$~cm$^{-2}$ are high even for an intrinsic AGN outflow (Pounds
et al 2003, hereafter P03; Reeves et al 2003). If these are instead of
local origin then it requires a tremendously significant change to our
understanding of the Galactic halo environment (McKernan et al 2004;
2005). 

Here we use new Suzaku data to show that the highly ionised
Fe absorption in PG~1211+143 is variable on a timescale of years. 
This conclusively demonstrates that
it is intrinsic to the AGN, and not associated with our Galaxy or
Local Group. We collate recent results to show that this is also
the case for most other AGN with strong Fe K absorption lines, showing
that powerful outflows are associated with luminous accretion flows. 

\section{The ionised absorption in PG~1211+143}

Chandra and XMM--Newton gratings gave the first high resolution X--ray
spectra of the warm absorbers in AGN (e.g. Blustin et al 2005,
McKernan, Yaqoob \& Reynolds 2007). However, only the Chandra HETGS
can cover the Fe K region, and this has very small effective
area. Thus most of the information on these absorption lines are from
the large effective area, moderate resolution CCD's on XMM-Newton.
These showed that some powerful AGN have strong absorption lines at
7~keV (observed frame) associated with large columns of highly ionised iron
(PG~1211+143: P03; Pounds \& Page 2006, hereafter PP06; PDS456: Reeves
et al 2003).

The results on PG~1211+143, at a redshift of $z=0.0809$, were
initially controversial. P03 claimed that the dip in
the spectrum at $\sim 7$~keV was an absorption line, which they
initially identified H--like Fe K$\alpha$ (6.97 keV,
lab--frame). Given the source redshift this requires an outflow
velocity of $\sim 24,000$~km/s i.e. 0.08c assuming an origin in the
AGN. Other, weaker but still significant absorption lines in the CCD
spectra at 2.68 and 1.47~keV could also be interpreted as arising from
material with the same outflow velocity if these are from H--like S
and Mg K$\alpha$, respectively.  (P03, PP06). 
The higher resolution RGS instrument also shows lines from
lower Z elements at low energies. P03 identified
these with transitions which required similarly high velocities as the
high ionisation material, while Kaspi \& Behar (2006) used different
identifications to obtain a lower velocity outflow solution for these
lines.  

Nonetheless, the origin of the RGS features is not the major
issue. Only the CCD's on XMM-Newton can show the existence of material
which is so highly ionised that heavy elements (S, Mg, Fe) are
predominantly He-- or H--like.  Surprisingly, Kaspi
\& Behar (2006) identified the S and Mg K-shell lines with weaker  
He-like K$\beta$ lines (with little velocity offset), instead of 
the blue-shifted H-like K$\alpha$ transitions, 
which begs the question as to why the corresponding stronger K$\alpha$ 
lines are not seen. These authors also model the 
Fe K absorption in the 2001 XMM-Newton data with an edge at 7.27~keV
(rest frame) of optical depth $\tau\sim0.5$,  corresponding 
to a large column ($N_{\rm H}>10^{23}$\,cm$^{-2}$)
of very low ionisation state matter (Fe VII/VIII); this should have
substantial low energy absorption which is clearly not present in the
data. 
As we show below, accounting for any low ionisation 
absorption (via partial covering) in a self consistent 
photoionisation model does not remove the 
requirement for the blue-shifted highly ionised iron K absorber.

The only
self--consistent option for the features around 7~keV in the
XMM-Newton data is K$\alpha$ absorption lines from H-- or He--like iron. This
requires a large blueshift (0.08c or 0.14c respectively) if the
absorption is intrinsic to the AGN, with the implication being that
the kinetic energy associated with this material can dominate the AGN
energetics (Pounds \& Reeves 2007).

\section{Data Reduction}

\subsection{Suzaku Analysis}

\PG\ was observed by Suzaku (Mitsuda et al. 2007) between 24-27
November 2005.  In this paper we discuss data taken with the 4 XIS
(X-ray Imaging Spectrometer; Koyama et al. 2007) CCDs. 
Events files from official version 2.0 of the Suzaku pipeline processing were used.
The \PG\ Suzaku event files were screened using identical criteria to those 
described in Reeves et al. (2007) for the Suzaku observation of MCG\,-5-23-16. 

Subsequently source spectra from the XIS CCDs were extracted from circular regions
of 2\arcm\ radius centred on the source, in the off-axis
HXD nominal pointing position. Background spectra were
extracted from 4\arcm\ circles offset from the source region, avoiding
the calibration sources on the corners of the CCD chips.  XIS response
files (rmfs) and ancillary response files (arfs) were generated using
the \textsc{xisrmfgen} and \textsc{xissimarfgen} scripts including
correction for the hydrocarbon contamination on the optical blocking
filter (Ishisaki et al. 2007).  A net XIS source exposure of 97.5 \,ks
was obtained for each of the 4 XIS chips.  The 3 front illuminated
XIS chips (XIS 0,2,3) were used in this paper, as they have the greatest 
sensitivity at iron K. These chips were found to produce consistent spectra within
the statistical errors, so the spectra and responses were combined to
maximise signal to noise. The net source count rate for the 3 XIS
combined was $0.7984\pm0.003$\,counts\,s$^{-1}$, with background only
3\% of the source rate.  The XIS source spectrum was binned to a
minimum of 100 counts per bin to enable the use of $\chi^{2}$
minimisation.  Errors are quoted to
90\% confidence for 1 parameter (i.e. $\Delta\chi^{2}=2.7$).  

\begin{figure}
\begin{center}
\rotatebox{-90}{\includegraphics[width=6cm]{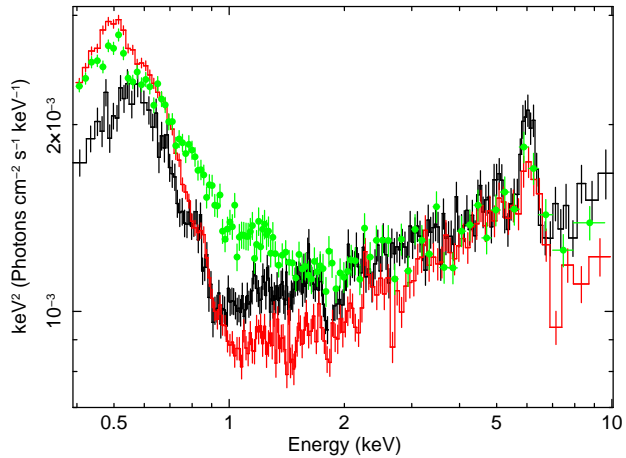}}
\end{center}
\caption{Fluxed spectra from the 2001 XMM-Newton (red crosses), 2004 XMM-Newton 
(green circles) and 2005 Suzaku (black crosses) 
observations of \PG, compared to a $\Gamma=2$ continuum. The spectra 
are plotted in the observed frame (at $z=0$). The downturn at 
lower energies is due to the Galactic column of $2.8\times10^{20}$\,cm$^{-2}$. 
Strong soft X-ray excesses below 1 keV are observed in all 3 spectra, 
which are most prominent in the 2001 and 2005 observations. The deep 
absorption feature at 7 keV is detected in the 2001 XMM-Newton observation, 
but not in the 2004 XMM-Newton and 2005 Suzaku observations. The spectra have been re-binned by a 
factor $\times2$ for clarity.}
\end{figure}

\subsection{XMM-Newton Analysis}

XMM-Newton observed \PG\ twice, on 2001 June 15 for 53\,ks and 2004
June 21 for 57\,ks but this second exposure was affected by high
particle background, resulting in only 25\,ks of usable data.  The
EPIC-pn spectra used in this paper are identical to those described in
PP06 and Pounds \& Reeves (2007), which were processed using the XMM-Newton SAS v6.5 
software. The EPIC-MOS spectra are
consistent with the EPIC-pn, with the same 7 keV absorption feature
present in the iron K band, as discussed by PP06.

\section{A Comparison Between Suzaku and XMM-Newton Observations}

The 2001--2005 spectra from 0.4-10 keV are plotted in Figure 1,
relative to a $\Gamma=2$ power-law continuum. The CCD's on both
XMM-Newton and Suzaku have similar responses so the data are directly
comparable. A Galactic column of $2.8\times10^{20}$\,cm$^{-2}$ is
included in all the spectral fits. From the spectra, it is clear that
there is little change in the overall spectral shape or
normalisation. All 3 observations show a strong soft X-ray excess
below 1 keV, though it is marginally more prominent during the 2001
XMM-Newton and 2005 Suzaku observations. However the deep iron K
absorption feature observed at 7 keV in 2001 (Pounds et al. 2003)
which is readily apparent even on this broad band spectral plot, is
not present at the same strength in either the 2004 or 2005
data. This feature has varied significantly, ruling out a
local origin for the absorber.

To quantify this, we use only the 2001 XMM-Newton and 2005 Suzaku
observations as these have the best signal--to--noise. We fit the
datasets simultaneously over the 2--10\,keV bandpass to avoid the
complexity associated with the soft X--ray excess. A power law with
fixed Galactic absorption gives a very similar index between the two
datasets as the continua are very similar (see Fig 1), but the
overall fit is poor ($\chi^{2}/{\rm dof}=374/154$ and $369/249$ for
the XMM-Newton and Suzaku data, respectively).  Fig. 2 shows the ratio
of the data to the model for this fit, 
where as well as the obvious absorption feature in the
XMM-Newton data, it is clear that both spectra have iron emission at
$\sim 6.5$~keV.  Adding a Gaussian line (with intrinsic width
constrained to be equal across the two datasets) gives a line energy
and equivalent width consistent between the two datasets, at $6.5$~keV
and $250$~eV respectively. The intrinsic line width is resolved with a 
FWHM velocity of $25000$\,km\,s$^{-1}$ i.e. typical of 
radii \ltsima$100R_{g}$ from the black hole.

\begin{table*}
\begin{tabular}{cccccccccc}
  \hline
Model &  $\Gamma^{a}$  & Norm $\times 10^{-3\,a}$ & line E (keV)$^{b}$ & $\sigma$ (keV)$^{b}$ 
& EW (eV)$^{b}$ &  line E (keV)$^{c}$ & $\sigma$ (keV)$^{c}$ &
  EW (eV)$^{c}$ & $\chi^2_\nu$ \\
       &   &      &              &          &    &  $N_{\rm H}$~cm$^{-2}$ & $\log\xi$ &
  Cov Frac &  \\
  \hline

po &$1.77\pm 0.03$ & 0.88 \\
   & $1.76\pm 0.03$ & 0.99 &&&&&&& 744/403\\

Gau & $1.93\pm 0.05$ & 0.96 & $6.63\pm 0.09$ & $0.24\pm 0.04$ & $250^{+80}_{-70}$ & $4.36^{+0.23}_{-0.29}$ & $1.07^{+0.24}_{-0.20}$ & $810^{+210}_{-180}$ \\
& $1.87\pm 0.04$ & 1.07 & $6.52^{+0.04}_{-0.05}$ & & $270\pm 50$ & 
$4.39^{+0.38}_{-0.48}$ & & $290\pm 100$ & 441.4/393 \\

Pcfxi & $2.50^{+0.16}_{-0.17}$ & $4.07$ & $6.55^{+0.09}_{-0.07}$ &
  $0.20\pm  0.04$  & $235\pm 50$    & $20^{+8}_{-4}$ & 
$2.18^{+0.29}_{-1.20}$ &   $0.78^{+0.08}_{-0.08}$  \\

& $2.10^{+0.12}_{-0.10}$ & 2.00 & $6.53^{+0.05}_{-0.02}$ & & $210\pm 45$ & 
$16^{+12}_{-6}$ & & $0.42^{+0.09}_{-0.12}$ & 439.6/393 \\

  \hline

\end{tabular}

\caption{Fit parameters and uncertainties ($\Delta \chi^2=2.7$) for
the XMM-Newton 2001 and Suzaku 2005 spectra from PG~1211+143 for the models
shown in Fig 3.$^{a}$ Continuum photon index and normalisation (units 
photons\,cm$^{-2}$\,s$^{-1}$\,keV$^{-1}$ at 1\,keV). $^{b}$ Energy (keV), intrinsic 
width $\sigma$ (keV) and equivalent width (eV) of the iron line, 
quoted in the quasar rest frame. $^{c}$ Energy (keV), intrinsic 
width $\sigma$ (keV) and equivalent width (eV) of the broad redshifted 
Gaussian (Gau model), or for the partial covering (Pcfxi) 
model, column density $N_{\rm H}$ (units $\times10^{22}$\,cm$^{-2}$), 
ionisation parameter $\xi$ (units erg\,cm\,s$^{-1}$) and covering fraction.} 
\end{table*}

However, there is still significant curvature in the spectrum, as
shown by the poor combined $\chi^2_\nu=582/398$.  This can be
phenomenologically modelled by an additional Gaussian line, again with
intrinsic width tied between the two datasets. This gives an
acceptable fit, with $\chi^2=441.3/393$, for a very broad line, with
an intrinsic width of $1.1\pm 0.2$~keV. The line energy is consistent
between the two datasets at $\sim 4.4$~keV but its equivalent width is
much larger in the XMM-Newton data than in the Suzaku data, at $\sim
800$~eV compared to $\sim 290$~eV (see Table 1), showing that there is
more curvature in the XMM-Newton data. Since it is also the XMM-Newton
data which show the line absorption feature, it is possible that 
this curvature is associated with the absorber, rather than with a
broad, redshifted iron line from reflection from the very 
innermost (i.e. $<10{\rm R}_{\rm g}$) accretion disc. 

We phenomenologically model this by partial covering of the source by
partially ionised material (e.g. Miller et al., 2007; Turner et al.,
2007), using a grid of XSTAR (v2.1l) photoionisation
models based upon the publicly available
tabulated 'grid 25'. This uses the Fe K treatment of Kallman et al. (2004), 
a turbulent velocity of
$\sigma=200$\,km\,s$^{-1}$ and solar abundances (Grevesse \& Sauval 1998),
while the illuminating continuum from 1--1000 Rydbergs has
$\Gamma=2.2$ rather than $\Gamma=2$, and spans a wider (but coarser)
range of column (0.05 to 500$\times 10^{22}$ cm$^{-2}$, 10 points) and
$\log\xi$ (-3 to 6, 12 points). This partially ionised absorption is
applied to a fraction $f$ of the assumed continuum, while the
remaining ($1-f$) is not affected by this material.
\footnote{This model is publicly available as an additional model for
XSPEC from {http://heasarc.gsfc.nasa.gov/docs/xanadu/xspec/newmodels.html}}


We fix the redshift of the partial coverer to that of
the AGN, and tie the ionisation parameter of
this material (but not its covering fraction) between the two
datasets. This model gives an equally good fit
as the broad Gaussian model, with $\chi^2=439/393$ for $\log \xi \sim
1.2$. The column of the ionised material is consistent between the two
datasets at $\sim 2\times 10^{23}$~cm$^{-2}$, but the covering
fraction is much larger in the XMM-Newton spectrum (80 per cent
compared to 40 per cent), showing that the spectral curvature is
more evident in these data.  We use these two different models
(hereafter termed Gau and Pcfxi, respectively) to assess the
robustness of the change in the narrow absorption features to
different continuum placement. Note the partial covering model is 
insensitive to the turbulence assumed, as its effect from 2--10\,keV 
is to reproduce the continuum curvature in the spectrum, from bound--free absorption 
such as the iron K-shell edge.

\begin{figure}
\begin{center}
\rotatebox{-90}{\includegraphics[width=6cm]{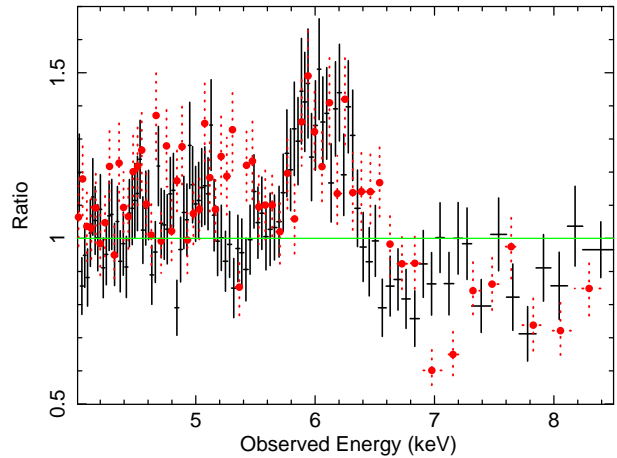}}
\end{center}
\caption{Data/Model ratio residuals at iron K to a simple power-law
fit to the 2--10\,keV spectrum of \PG. The 2001 XMM-Newton observation
is in red, 2005 Suzaku in black.  The absorption feature at 7 keV (7.6
keV QSO rest frame) is apparent in the XMM-Newton data, but
not in the Suzaku observations, showing that the iron K absorption has
varied over 4 years. Broad iron
K$\alpha$ emission is present in all observations at 6 keV (6.5 keV
rest frame), as well as curvature at 4-6~keV which can be
modelled either by a very broad, redshifted line or by continuum
complexity.}
\end{figure}

Fig 3 shows the $\chi^2$ residuals to the power-law, 
Gau and Pcfxi models, respectively. The absorption line is
still present in the XMM-Newton data, even against the Pcfxi model which
contains the iron K-shell edge and its associated absorption line 
structure (Figure 13, Kallman et al. 2004). We fit
an additional inverted Gaussian of fixed width ($\sigma=0.1$\,keV) to the
XMM-Newton data and find that the fit is significantly improved, to
$\chi^2_\nu=418.7/391$ and $417.6/391$ for the Gau and Pcfxi continuum
models, respectively. The line energy and equivalent width in both is
$\sim 7.6$\,keV ($7.05\pm0.05$\,keV observed) and
$110^{+45}_{-40}$~eV, respectively.  We allow a line with the same
energy and width in the Suzaku data, and find a 90\%
confidence upper limit to its equivalent width of $<25$~eV.

\begin{figure}
\begin{center}
\rotatebox{0}{\includegraphics[width=0.45\textwidth]{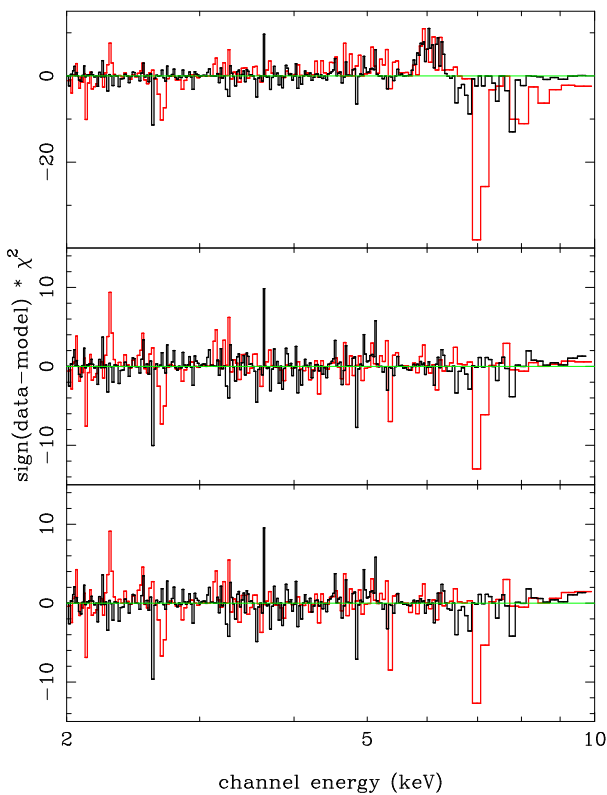}}
\end{center}
\caption{$\Delta\chi^{2}$ residuals against
a power law (top), iron emission line and broad redshifted Gaussian
(middle) and iron emission line and partial covering by partially
ionised material (bottom) for the XMM-Newton 2001 (red) and Suzaku
2005 (black) data, respectively. 
The absorption feature at $\sim 7$~keV 
in the XMM-Newton data is significantly stronger than in the
Suzaku data irrespective of the phenomenological continuum model.}
\end{figure}




\section{The Origin of the Iron K shell Absorption}

Thus the absorption line is significantly weaker in
the 2005 Suzaku data, so the absorption has changed on a timescale of
$<4$ years. Either the column has physically moved, or its ionisation
state has changed (or both). The former puts a limit on the size scale
of 4 light years, unfeasibly small for any diffuse
Galactic halo or intergalactic emission (which should be on kpc
scales), while the latter sets a limit on the density through the
recombination timescale, $\tau_r=1/(\alpha_r(T) n_e)$, where $\alpha_r$
is the recombination rate, which is dependent on the temperature, T.
For a WHIM origin, the absorption line must be associated with Fe XXVI
in order that its energy matches that observed. Such material should
also be collisionally ionised, and requires temperatures $>10^{7.5}$~K in
order to be the dominant ion. This gives a recombination 
rate of $\alpha_r\sim 2.5\times 10^{-12}$ cm$^{3}$ s$^{-1}$
(e.g. Arnaud \& Rothenflug 1985), so for this to be less than 4 years
requires densities $> 4\times 10^3$~cm$^{-3}$, which is 6 orders of magnitude
above the typical density even for a local halo component
(e.g. Bregman \& Lloyd--Davies 2007).

We model the Fe K absorption line using the {\sc warmabs} XSTAR
photo-ionisation code incorporated into {\sc xspec}, adopting 
solar abundances. We assume a turbulent velocity of 5000 km s$^{-1}$, below the upper limit of
$<8000$ km s$^{-1}$ set by the observed intrinsic width of the
line of $\sigma<0.2$~keV.
This gives a column of $8\pm3\times10^{22}$\,cm$^{-2}$, 
with an ionisation parameter of $\log\xi=2.8^{+0.2}_{-0.3}$ 
at an observed redshift of $z=-0.07\pm0.01$ i.e.
implying an AGN outflow velocity of $\sim0.15$c.
While the derived column is
dependent on the turbulent velocity, the curve of growth is in the
linear regime here, so increasing the turbulent velocity does not
increase the equivalent width of the line, leading to a robust lower limit
of $N_{\rm H}>5\times10^{22}$\,cm$^{-2}$.

The lower limit on the density, combined with the column density, can give an upper limit on the 
size scale $\Delta R$ of the absorber as $N_H=n\Delta R$. 
A huge column of $N_H\ge 10^{25}$ cm$^{-2}$ is required in order to produce
an iron absorption line
with equivalent width of 100 eV if the gas velocity is simply set by its
temperature of $10^{7.5}$~K (Kotani et al. 2000; 2006) as expected 
from any local hot gas. 
Such a large column is inconsistent with the spectrum, as this 
produces a strong dip at 8--10~keV (rest frame) due to the Fe K-shell edge structure 
(Kallman et al 2004), which is not present in the
data. Indeed from the lack of an iron K edge in 
PG\,1211+143, this sets an upper limit to the column of $<10^{24}$\,cm$^{-2}$
and requires that there must be substantial turbulent velocity ($>1000$\,km\,s$^{-1}$ 
derived from the curve of growth) in excess of the
thermal motions to model the line EW. This itself is inconsistent with a WHIM origin. 
With this, the upper limit on the absorber size scale
is $\Delta R < N_H/n =10^{24}/4\times 10^{3} < 2.5\times 10^{20}$~cm. 
This is unfeasibly
low to be associated with even the smallest local
diffuse hot gas, the kpc scale Galactic halo. 

Another constraint on the properties of the material comes from the emission 
measure $EM = n^2 V$ where $V=4\pi R^2 \Delta R$ is the volume, assuming that 
the gas is distributed evenly in a spherical shell at distance $R$. The total 
bremsstrahlung luminosity is $L_{brems}\sim 2\times 10^{-27} T^{1/2} EM$, 
so the surface brightness in solid angle $d\Omega$ is $L_{brems} /(4\pi R^2) 
(d\Omega/4\pi)$. For a solid angle of 1 arcmin$^2 = 10^{-7}$~sr this gives 
$\sim 1.3\times 10^{-35} T^{1/2} n N_H > 10^{-5}$ ergs s$^{-1}$ cm$^{-2}$ arcmin$^{-2}$, 
i.e. brighter than the Crab in every square arcmin in X-rays.  
Indeed the observed upper limit to the diffuse
sky surface brightness in the 3/4~keV band is $\sim 2\times 10^{-4}$
counts s$^{-1}$ arcmin$^{-2}$ (Bregman \& Lloyd--Davies 2007),
corresponding to a flux of $<2\times 10^{-15}$ ergs s$^{-1}$ cm$^{-2}$
arcmin$^{-2}$, 10 orders of magnitude below the predicted value. 

Thus the variability and column appear to rule out any sort of
non-AGN origin for the absorbing material. The best fit identification
by PP06 of the highly ionised absorption as being dominated by He--like Fe
also rules out a local origin as the line is
then not close to the rest wavelength of the transition, showing
a blueshift with respect to the local frame of $-0.07$. This is 
because the Mg, Si and S K-shell absorption seen in the XMM-Newton CCD spectrum 
can be fit with the same layer of absorption if the ionisation state 
of iron is lower than H-like. This
mismatch between the outflow velocity of the absorber and recessional
velocity of the AGN is common.  There are now many other AGN
with blueshifted Fe K$\alpha$ indicating an outflow, and we collate
these from the literature in Fig. 4. Aside from the 3 low velocity examples 
in Figure 4, none of the AGN outflow velocities coincide with the recession velocity. 
In IC~4329a and MCG\,-5-23-16 the absorption lines are observed at 
7.7~keV (Markowitz et al. 2006; Braito et al. 2007)
inconsistent with the rest energy of any Fe K$\alpha$
transition line. A further example is from 
the new 2007 Suzaku data from PDS456 (Reeves et al. 2007, in prep),  
which shows iron K absorption lines observed at 7.6 and 8.2~keV, inconsistent 
with local absorption from iron K$\alpha$.  
Indeed the iron K absorption features seen in high redshift 
BAL QSO's are likewise very far from the iron K$\alpha$ rest energies for local
material (Chartas et al 2002, 2003).

\begin{figure}
\begin{center}
\rotatebox{-90}{\includegraphics[width=6cm]{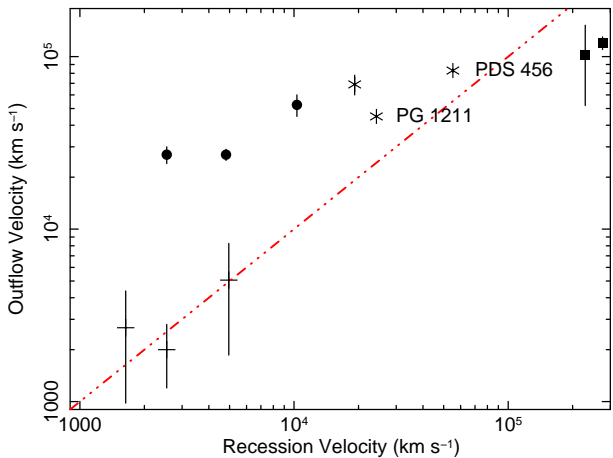}}
\end{center}
\caption{The inferred outflow velocity for the Fe K$\alpha$ absorption
lines versus the host galaxy recession velocity. Only the three lowest
velocity points (crosses) are kinematically consistent 
(within uncertainties) with a local rest frame energy 
(left to right; NGC 1365, Risaliti et al. 2005; MCG\,-6-30-15, 
Young et al. 2005; IRAS\,13197-1627, Miniutti et al. 2007). 
The filled circles represent the blue-shifted Fe K absorbers in the Seyfert 
galaxies MCG\,-5-23-16 (Braito et al. 2007), IC 4329a (Markowitz et al. 
2006) and Markarian 509 (Dadina et al. 2005). Stars represent absorption 
systems observed in nearby quasars (PG\, 1211+143, this paper; 
PG\,0844+349, Pounds et al. 2003b; PDS 456, Reeves et al. 2007, in prep). The filled 
squares correspond to two BAL quasars, PG 1115+080 (Chartas et al. 2003) 
and APM\,08279+5255 (Chartas et al. 2002).} 
\end{figure}

\section{Conclusions}

We show that the observed variability in the absorption line in
PG~1211+143 between the XMM-Newton and Suzaku data taken 4 years apart
conclusively rules out a diffuse gas origin such as the local Galactic
halo or WHIM. The gas must be associated with the AGN, as is further
evidenced by its large column density in the XMM-Newton 2001 data which
is far too high for local or intergalactic gas. The conclusive identification
with the AGN, and the implausibility of any alternative line
transition other than iron K$\alpha$ means that the large outflow
velocity of $\sim 0.1$c is inescapable. Such material is 
predicted from the winds which are produced from luminous
accretion discs in AGN (Proga \& Kallman 2004).  

The coincidence of the outflow velocity with the source redshift in
these data is not significant. Other powerful AGN which show iron K
absorption features clearly show a range of (intrinsic and observed)
velocities, removing the apparent trend for the line energy to appear
at the rest energy for these transitions. Thus it is clear that this
material is a mildly relativistic outflow from the AGN. Its high
velocity means that its kinetic energy can be comparable to the
bolometric radiated luminosity of the AGN (King \& Pounds 2003; Pounds
et al 2003; Pounds \& Reeves 2007), yet its high ionisation means that
it is effectively invisible in all other wavebands. While such a
major component of AGN energetics could go unnoticed in individual
objects, there is increasing evidence that strong AGN feedback
controls galaxy formation and evolution. Mildly relativistic winds
provide more efficient heating than the jet due to their impact on a
larger area, so these highly ionised disc winds may be the key to
understanding the growth of structure in the Universe.

\end{document}